\documentclass[a4paper]{article}

\advance\voffset by -20mm
\advance\hoffset by -20mm
\usepackage[dvips]{graphicx}
\usepackage{pstricks}
\usepackage{pst-plot}
\usepackage{amsfonts}

\title{Spatial Coupling of a Lattice Boltzmann fluid model with a Finite Difference Navier-Stokes solver}
\author{Jonas~Latt\footnote{Computer Science Department, University of Geneva, 1211 Geneva 4, Switzerland}
        \and Bastien~Chopard
        \and Paul~Albuquerque}

\begin{document}
\maketitle
\begin{abstract}
In multiscale, multi-physics applications, there is an increasing need for coupling numerical solvers that are each applied to a different part of the problem. Here we consider the case of coupling a Lattice Boltzmann fluid model and a Finite Difference Navier-Stokes solver. The coupling is implemented so that the entire computational domain can be divided in two regions, with the FD solver running on one of them and the LB one on the other.

We show how the various physical quantities of the two approaches should be related to ensure a smooth transition at the interface between the regions. We demonstrate the feasibility of the method on the Poiseuille flow, where the LB and FD schemes are used on adjacent sub-domains.

The same idea can be also developed to couple LB models with Finite Volumes, or Finite Elements calculations.

The motivation for developing such a type of coupling is that, depending on the geometry of the flow, one technique can be more efficient, less memory consuming, or physically more appropriate than the other in some regions ({\it e.g.} near the boundaries), whereas the converse is true for other parts of the same system. We can also imagine that a given system solved, say by FD, can be augmented in some spatial regions with a new physical process that is better treated by a LB model. Our approach allows us to only modify the concerned region without altering the rest of the computation.
\end{abstract}

\section{Introduction}\label{intro}
When it comes to the numerical analysis of fluid flows, one has the choice between many different models. On one hand there are solvers based on the discretization of the Navier-Stokes equations, for example by a numerical scheme of finite differences, finite elements or finite volumes. On the other hand, a new category of solvers have emerged over the past decades that are based on kinetic theories. They describe the fluid dynamics at a molecular level and can be seen as discretization schemes for the Boltzmann equation.

In this paper we consider the possibility of solving separate spatial regions of a simulations with a different solver. In particular, we are interested in coupling a finite difference (FD) solver of the two-dimensional Navier-Stokes equations with a lattice Boltzmann (LB) method, {\it i.e.} a solver for the Boltzmann equation. The motivation is that depending on the nature of the region, optimal efficiency may be reached with a different solver. The same reasoning applies to the implementation of boundary conditions that are more or less naturaly formulated in a given numerical scheme.
 
As an example, consider the computation of the drag force experienced by a rigid body placed in a uniform stream. During the numerical treatment of this problem, two kinds of boundary conditions need to be implemented, one for the boundaries of the body and one for the boundaries of the computational domain.
In the first case, the emphasis is put on a correct representation of physical properties of the body. This part of the domain might therefore preferrably be solved with a LB model, which delivers a microscopic description of the physics. Furthermore, the LB model gives a direct access to physical quantities such as the drag force.
On the other hand, the boundaries of the computational domain are required to simulate a space that extends infintely in all directions. A lot of computational space can be saved if those boundaries implement an appropriate velocity field~(\cite{wittwer:03}). Indeed, the velocity field can be shown to be independent of the geometry of the body at a large enough distance. It depends only on the drag force and can be computed analytically. The implementation of this analytical solution on the boundaries finds a very natural formulation in the FD scheme, which is based on macroscopic variables.

In order to couple a LB with a FD model, it is crucial to understand how the LB set of variables is related to the FD set and vice versa. Our method follows the same arguments as the ones developed in~\cite{davide:04}, where the coupling between a LB and a FD solver for the heat equation is presented. The connection between the two models is achieved through a first-order expansion of the LB variables around a local equilibrium term. One finds that the zeroth-order terms of the expansion are related to the macroscopic quantities ({\it i.e.}the FD variables), whereas the first-order terms depend on gradients of those quantities.

The paper is organized as follows. Section~\ref{models} gives a brief overview of the chosen LB and FD models. Conceptual differences between the models are pointed out in view of the formulation of the coupling algorithm which is develoiped in section~\ref{coupling}. This rather technical section contains the first order expansion of the LB variables and a guideline to the LB/FD coupling. The algorithm is validated in section~\ref{poiseuille} on a Poiseuille flow and is found to be of second order in the velocities. Section~\ref{conclusion} draws a conclusion and presents some plans for future work.

\section{The numerical models}\label{models}
The FD model used in our study implements a finite difference scheme on a staggered grid. It is explicit in the velocities and implicit in the pressure. The chosen LB model implements the LBGK formulation, in which the dynamics are expressed in form of a relaxation towards a local equilibrium term.

In the following, we will restrict the discussion to an overview and a couple of technical aspects of those two models. The interested reader will find more details on the LB model in the references~\cite{succi-book:01}, \cite{BC-livre}~and~\cite{wolf-gladrow:00}. The FD model is explained in many details in the reference~\cite{griebel-book}. This book offers among others a reference to a complete implementation of the FD code in the C language.

\subsection{Spatial discretization}
We consider a two-dimensional, rectangular region $$\Omega=[0,l]\times[0,h]\in\mathbb{R}^2$$ on which we introduce a grid. This grid is divided into $i_{\mathrm{max}}$ cells of equal size in the $x$-direction and $j_{\mathrm{max}}$ cells in the $y$-direction, resulting in grid lines spaced at a distance $$\delta x = l/i_{\mathrm{max}}\quad\mathrm{and}\quad\delta y = h/j_{\mathrm{max}}$$.

\begin{figure}
\begin{center}
\caption{Choice of indices for FD and LB variables on a chosen grid cell $(i,j)$.}\label{cell}
\begin{pspicture}(10,6)
\psset{unit=0.5}
\psline[linewidth=0.05](4,5)(12,5)
\psline[linewidth=0.05](4,9)(12,9)
\psline[linewidth=0.05](6,3)(6,11)
\psline[linewidth=0.05](10,3)(10,11)
\psdots[dotstyle=*](8,5)(6,7)(8,9)(10,7)(8,7)
\psdots[dotstyle=square*,dotsize=0.4](6,5)(6,9)(10,5)(10,9)

\psline[linewidth=0.05]{->}(8,1)(8,4)
\psline[linewidth=0.05]{->}(2,7)(5,7)
\rput[bl](2.2,7.2){$j$}
\rput[bl](8.2,1.2){$i$}

\rput[b](6.1,5.1){$\vec f_{i-1,j-1}$}
\rput[bl](10.1,5.1){$\vec f_{i,j-1}$}
\rput[bl](10.1,9.1){$\vec f_{i,j}$}
\rput[b](6.1,9.1){$\vec f_{i-1,j}$}

\rput[bl](8.1,5.1){$v_{i,j-1}$}
\rput[bl](8.1,9.1){$v_{i,j}$}
\rput[bl](10.1,7.1){$u_{i,j}$}
\rput[bl](8.1,7.1){$p_{i,j}$}
\rput[b](6.1,7.1){$u_{i-1,j}$}

\psdots[dotstyle=*](14,4.3)
\psdots[dotstyle=square*,dotsize=0.4](14,3.3)
\rput[bl](14.5,4){FD variables}
\rput[bl](14.5,3){LB variables}

\end{pspicture}

\end{center}
\end{figure}
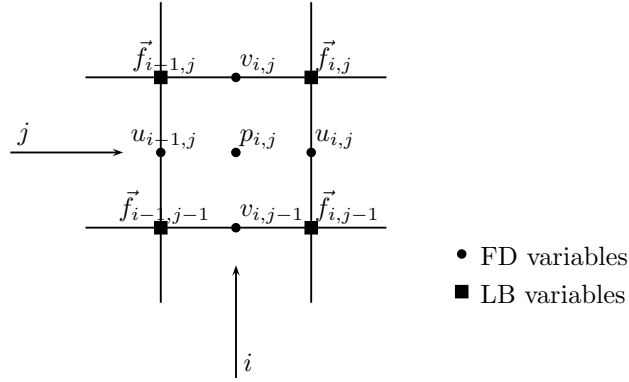

The FD model is based on three quantities that are defined on each cell: the pressure~($p$), the $x$-component~($u$) and the $y$-component~($v$) of the velocity. They are however placed on a staggered grid. A given index $(i,j)$ of the cell is assigned to the pressure at the cell center, to the $x$-component of the velocity at the right edge and the $y$-component at the upper edge (cf. Figure~\ref{cell}). The reason for this staggered arrangement is that it prevents possible pressure oscillations which could occur had all three variables $u$, $v$ and $p$ be evaluated at the same grid points.

The LB model uses nine variables $f_k, k=0\cdots8$ which are all evaluated at the same location of a cell. We fixed our choice on the upper right corner. This is to ensure that the LB and the FD model have a compatible interpretation of the location of the domain boundary $\delta\Omega$. Indeed, this boundary is defined on a cell edge in the FD model. Considering that most implementations of LB boundary conditions set the domain boundary on top of a LB node, this leads us to placing the LB node on the intersection of two cell edges.

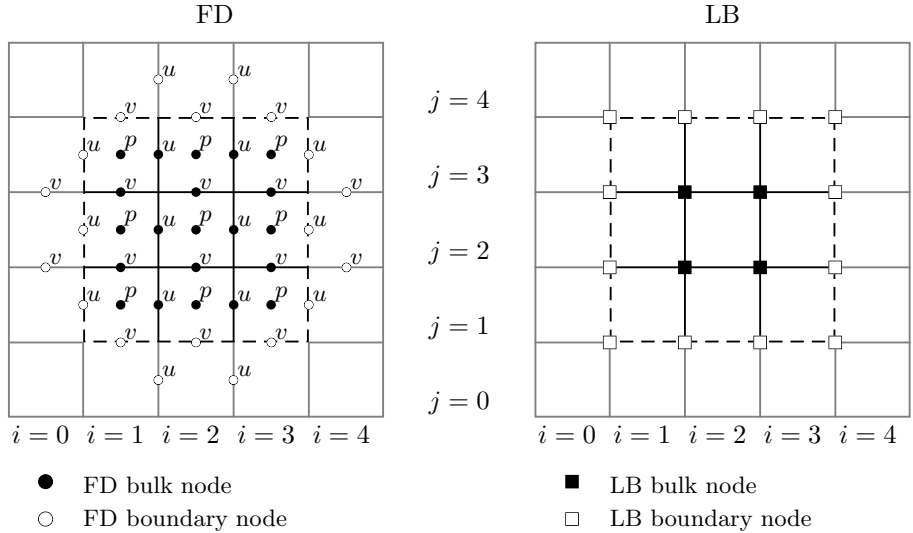
\begin{figure}
\begin{center}
\caption{Computational grid representing a domain $\Omega$ of size $(i_{\mathrm{max}}\cdot\delta x)\times(j_{\mathrm{max}}\cdot\delta y)$ with $i_{\mathrm{max}}=j_{\mathrm{max}}=3$. The left hand side depicts the staggered arrangement of the variables over the grid when the domain is resolved by a FD scheme. In the case of a LB solver, all variables are located on cell edges, as shown on the right hand side. The location of the boundary strip is indicated by a dashed line.}\label{grid}
\begin{pspicture}(12,8)
\psset{unit=0.5}
\psline[linewidth=0.05](2,8)(8,8)
\psline[linewidth=0.05](2,10)(8,10)
\psline[linewidth=0.05](4,6)(4,12)
\psline[linewidth=0.05](6,6)(6,12)
\psframe[linewidth=0.05,linestyle=dashed](2,6)(8,12)
\psframe[linewidth=0.05,linecolor=gray](0,4)(10,14)
\multips(0,6)(0,2){4}{%
  \psline[linewidth=0.05,linecolor=gray](0,0)(2,0)%
  \psline[linewidth=0.05,linecolor=gray](8,0)(10,0)%
}
\multips(2,4)(2,0){4}{%
  \psline[linewidth=0.05,linecolor=gray](0,0)(0,2)%
  \psline[linewidth=0.05,linecolor=gray](0,8)(0,10)%
}

\rput[tl](0.1,3.8){$i=0$}
\rput[tl](2.1,3.8){$i=1$}
\rput[tl](4.1,3.8){$i=2$}
\rput[tl](6.1,3.8){$i=3$}
\rput[tl](8.1,3.8){$i=4$}

\rput[b](12,4.1){$j=0$}
\rput[b](12,6.1){$j=1$}
\rput[b](12,8.1){$j=2$}
\rput[b](12,10.1){$j=3$}
\rput[b](12,12.1){$j=4$}

\multips(3,7)(1,0){5}{%
  \multips(0,0)(0,2){3}{%
    \psdots[dotstyle=*](0,0)
  }%
}

\multips(3,8)(2,0){3}{%
  \multips(0,0)(0,2){2}{%
    \psdots[dotstyle=*](0,0)
  }%
}

\multirput[bl](2.1,7.1)(2,0){4}{%
  \multirput[bl](0,0)(0,2){3}{\small$u$}%
}
\multirput[bl](3.1,6.1)(2,0){3}{%
  \multirput[bl](0,0)(0,2){4}{\small$v$}%
}
\multirput[bl](3.1,7.1)(2,0){3}{%
  \multirput[bl](0,0)(0,2){3}{\small$p$}%
}

\multirput[bl](0,0)(2,0){3}{%
  \psdots[dotstyle=o](3,6)(3,12)
}

\multirput[bl](0,0)(0,2){3}{%
  \psdots[dotstyle=o](2,7)(8,7)
}

\multirput[bl](0,0)(2,0){2}{%
  \psdots[dotstyle=o](4,5)(4,13)
  \rput[bl](4.1,5.1){\small$u$}
  \rput[bl](4.1,13.1){\small$u$}
}

\multirput[bl](0,0)(0,2){2}{%
  \psdots[dotstyle=o](1,8)(9,8)
  \rput[bl](1.1,8.1){\small$v$}
  \rput[bl](9.1,8.1){\small$v$}
}

\rput[bl](5,14.5){FD}

\psdots[dotstyle=*,dotsize=0.4](1,2.3)
\rput[bl](2,2){\small FD bulk node}
\psdots[dotstyle=o,dotsize=0.4](1,1.3)
\rput[bl](2,1){\small FD boundary node}

\psline[linewidth=0.05](16,8)(22,8)
\psline[linewidth=0.05](16,10)(22,10)
\psline[linewidth=0.05](18,6)(18,12)
\psline[linewidth=0.05](20,6)(20,12)
\psframe[linewidth=0.05,linestyle=dashed](16,6)(22,12)
\psframe[linewidth=0.05,linecolor=gray](14,4)(24,14)

\multips(14,6)(0,2){4}{%
  \psline[linewidth=0.05,linecolor=gray](0,0)(2,0)%
  \psline[linewidth=0.05,linecolor=gray](8,0)(10,0)%
}
\multips(16,4)(2,0){4}{%
  \psline[linewidth=0.05,linecolor=gray](0,0)(0,2)%
  \psline[linewidth=0.05,linecolor=gray](0,8)(0,10)%
}

\rput[tl](14.1,3.8){$i=0$}
\rput[tl](16.1,3.8){$i=1$}
\rput[tl](18.1,3.8){$i=2$}
\rput[tl](20.1,3.8){$i=3$}
\rput[tl](22.1,3.8){$i=4$}

\multips(18,8)(2,0){2}{%
  \multips(0,0)(0,2){2}{%
    \psdots[dotstyle=square*,dotsize=0.4](0,0)
  }
}
\multips(16,4)(2,0){4}{
  \psdots[dotstyle=square,dotsize=0.4](0,2)(0,8)
}
\psdots[dotstyle=square,dotsize=0.4](16,8)(16,10)
\psdots[dotstyle=square,dotsize=0.4](22,8)(22,10)

\rput[b](19,14.5){LB}

\psdots[dotstyle=square*,dotsize=0.4](15,2.3)
\rput[bl](16,2){\small LB bulk node}
\psdots[dotstyle=square,dotsize=0.4](15,1.3)
\rput[bl](16,1){\small LB boundary node}
\end{pspicture}

\end{center}
\end{figure}

The situation is depicted on Figure~\ref{grid} for a system of extent $i_{\mathrm{max}}=j_{\mathrm{max}}=3$. It shows also that as a result of the staggered arrangement, not all extremal grid points of the FD set of variables come to lie on the domain boundary. For this reason, an extra boundary strip of grid cells is introduced, so that the boundary conditions may be found by linear interpolations between the nearest grid points on either side.

\subsection{The FD model}\label{models:fd}
The FD model is based on a discretization of the incompressible Navier-Stokes equations
\begin{equation}
\partial_t\vec u +\left(\vec u\cdot \vec\nabla\right)\vec u = -\vec\nabla p+\nu\Delta u + \vec F.\label{eq:navier-stokes}
\end{equation}
and the continuity equation
\begin{equation}
\mathrm{div}\,\vec u=0\label{eq:continuity}
\end{equation}
The computation of the successive iterations $\left(u^{(t)},v^{(t)},p^{(t)}\right)\Rightarrow\left(u^{(t+1)},v^{(t+1)},p^{(t+1)}\right)$ contains two distinct steps:
\begin{enumerate}
\item Resolution of the poisson equation to obtain the new pressure field. This computation utilizes the values of the pressure and the velocity at the time $t$: $\left(u^{(t)},v^{(t)},p^{(t)}\right)\Rightarrow\left(p^{(t+1)}\right)$. In presence of Dirichlet boundary conditions, this procedure has a unique solution (except for an integration constant). In particular, there is no need for knowing the value of the pressure on the boundary.
\item Computation of the new velocity field according to a finite difference scheme. It uses the pressure field at step $t+1$: $\left(u^{(t)},v^{(t)},p^{(t+1)}\right)\Rightarrow\left(u^{(t+1)},v^{(t+1)}\right)$.
\end{enumerate}

\subsection{The LB model}\label{models:lb}
The LB model can be interpreted as a discretization of the Boltzmann transport equation on the chosen lattice. The possible velocities for the pseudo-particles are the vectors $\vec v_k$. They are chosen so as to match the lattice directions: if $\vec r$ is a lattice site, $\vec r+\vec v_k\delta t$ is also a lattice site. In the present case, we use a so-called D2Q9 lattice with nine possible velocities: a zero velocity to describe the population of rest particles, four velocities for the horizontal and vertical directions, and four velocities for the diagonal directions. We restrict our considerations to a lattice with equal spacing in the $x$ and $y$ directions, {\it i.e.} $\delta x = \delta y =: \delta r$. The LB model can also be implemented on different kins of lattices, but it needs to satisfy a couple of isotropy requirements. For example, the second- and third-order tensors of the D2Q9 lattice are
\begin{eqnarray}
\sum_k m_kv_{k\alpha}v_{k\beta} &=& \frac{v^2}{b}\delta_{\alpha\beta}\quad\textrm{and}\label{eq:isotropy2}\\
\sum_k m_kv_{k\alpha}v_{k\beta}v_{k\gamma} &=& 0\label{eq:isotropy3},
\end{eqnarray}
where $\delta_{\alpha\beta}$ is the Kronecker symbol, and $b$ is a constant that is characteristic for the lattice.

The system is described by $9$ corresponding density distribution functions $f_k(\vec r,t), k=0\ldots8$, representing the distribution of particles entering site $\vec r$ at time $t$ and moving in direction $\vec v_k$. Physical quantities such as the local particle density and velocity are defined from moments of these distributions:
\begin{equation}\label{eq:macroscopics}
  \rho=\sum_{k=0}^8 m_k f_k\quad\textrm{and}\quad\vec u=\sum_{k=0}^8 m_k f_k\vec v_k.
\end{equation}
The $m_k$ are the lattice weights: those constant values compensate for the different lengths of diagonal and non-diagonal directions.

Although this fluid model is compressible (the density is space and time dependent), it can be shown to solve the incompressible Navier-Stokes equations in the  small Mach number regime. The fluid pressure is related to the fluid density by the ideal gas state equation $p=c_s^2\rho$, where $c_s^2$ is the speed of sound. Therefore, in the LB model, there is no need to solve the Poisson equation for the pressure. It is all built-in in the equation of motion for the $f_k$. However, there exists not always a straightforward way of treating the pressure on the boundaries, and an appropriate boundary condition has to be found.

In the BGK approximation, the particle collision is obtained by a relaxation with coefficient $\omega$ to a truncated Maxwell-Boltzmann equilibrium distribution function $f^{(\mathrm{eq})}$, which depends only on the local fluid density and velocity:
\begin{equation}\label{eq:lbgk}
f_k(\vec r+\vec v_i\delta t,t+\delta t)-f_k(\vec r,t)= -{\omega}\left(f_k(\vec r,t)-f_k^\mathrm{(eq)}(\vec r,t)\right) + \frac{b\,\delta t}{v^2} \vec F\cdot\vec v_i,
\end{equation}
where
\begin{equation}\label{eq:f_eq}
f_k^\mathrm{(eq)}(\vec r,t)=a\,\rho\left(1+\frac{b}{a}v_{i\alpha}u_\alpha+\frac{1}{2}\left(\frac{b}{a}v_{i\alpha}u_\alpha\right)^2-\frac{b}{2a}u^2\right).
\end{equation}
In these formulae and in the further developments, a repeated greek index inside a multiplicative term implies a sum on this index. The term $a$, like $b$, is a lattice constant.

One particularity of this model is that it satisfies the mass and momentum conservation at the molecular level:
\begin{eqnarray}
\sum_k m_k f_k\left(\vec r+\vec v_i\delta t,t+\delta t)-f_k(\vec r,t)\right) = 0\quad \textrm{and}\label{eq:lb-mass-conservation}\\
\sum_k m_k v_{k\alpha}\left(f_k(\vec r+\vec v_i\delta t,t+\delta t)-f_k(\vec r,t)\right) = \delta t\,F_{\alpha}.\label{eq:lb-momentum-conservation}
\end{eqnarray}

\section{The coupling algorithm}\label{coupling}
\subsection{The FD-LB interface}\label{coupling:interface}
In this section, we cut $\Omega$ into two subdomains $\Omega_1$ and $\Omega_2$ such that $\Omega=\Omega_1\bigcup\Omega_2$. We apply in $\Omega_1$ the FD method and in $\Omega_2$ the LB method respectively. For facility, we assume that the subdomains are rectangular, $\Omega_1$ occupying the left hand side and $\Omega_2$ the right hand side of the domain. This procedure can however be extended with few changes to a general boundary.

\begin{figure}
\begin{center}
\caption{Subdivision of the computational domain $\Omega$ into a FD subdomain ($\Omega_1$) and a LB subdomain ($\Omega_2$). One lattice cell on the interface between $\Omega_1$ and $\Omega_2$, at the position $i=i_\mathrm{int}$, is computed by both methods. The boundary nodes of the subdomains are represented by white symbols. They must be implemented by means of a coupling term between the two methods.}\label{fig:interface}
\begin{pspicture}(9,5)
\psset{unit=0.5}
\psline[linewidth=0.05,linecolor=gray](1,3)(17,3)
\psline[linewidth=0.05,linecolor=gray](1,5)(17,5)
\psline[linewidth=0.05,linecolor=gray](1,7)(17,7)
\multips(2,2)(2,0){8}{
  \psline[linewidth=0.05,linecolor=gray](0,0)(0,6)
}
\multirput(0,0)(2,0){3} {
  \multirput(0,0)(0,2){3} {
    \psdots[dotstyle=*](5,3)
    \rput[bl](5.1,3.1){$v$}
  }
  \multirput(0,0)(0,2){2} {
    \psdots[dotstyle=*](5,4)
    \rput[bl](5.1,4.1){$p$}
    \psdots[dotstyle=*](4,4)
    \rput[bl](4.1,4.1){$u$}
  }
}

\psdots[dotstyle=o](10,4)(10,6)(11,3)(11,5)(11,7)
\rput[bl](10.1,4.1){$u$}
\rput[bl](10.1,6.1){$u$}
\rput[bl](11.1,3.1){$v$}
\rput[bl](11.1,5.1){$v$}
\rput[bl](11.1,7.1){$v$}

\psdots[dotstyle=square*,dotsize=0.4](10,3)(10,5)(10,7)(12,3)(12,5)(12,7)
\psdots[dotstyle=square, dotsize=0.4](8,3)(8,5)(8,7)

\rput[b](5,8){FD}
\rput[b](13,8){LB}
\rput[b](3,5.2){$\Omega_1$}
\rput[b](15,5.2){$\Omega_2$}

\psline[linewidth=0.05,linestyle=dashed](8,1)(8,9)
\psline[linewidth=0.05,linestyle=dashed](10,1)(10,9)
\rput[b](9,9){\small$i=i_{\mathrm{int}}$}

\psdots[dotstyle=*,dotsize=0.4](1,1.3)
\rput[bl](2,1){\small FD bulk node}
\psdots[dotstyle=o,dotsize=0.4](1,0.3)
\rput[bl](2,0){\small FD boundary node}

\psdots[dotstyle=square*,dotsize=0.4](11,1.3)
\rput[bl](12,1){\small LB bulk node}
\psdots[dotstyle=square,dotsize=0.4](11,0.3)
\rput[bl](12,0){\small LB boundary node}

\end{pspicture}
\end{center}
\end{figure}
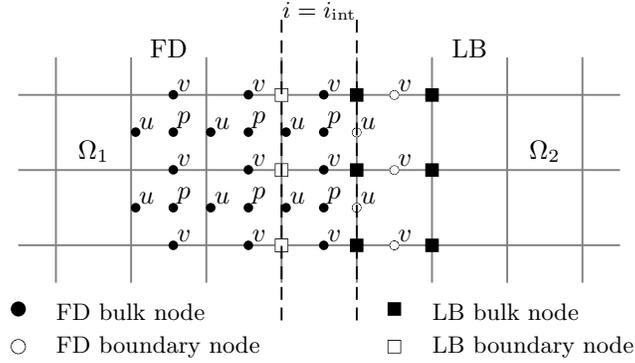

The way boundary conditions are implemented in the FD and the LB scheme has already been touched upon in section~\ref{intro}. In particular, Figure~\ref{grid} describes the position of the extremal grid points (drawn as white circles and squares) to which a boundary value must be furnished. On the coupling interface between $\Omega_1$ and $\Omega_2$, those boundary values are taken from the boundaries of the opposite domain. As a consequence of the staggered arrangement of the LB values with respect to the FD values, there is a need for an overlap between $\Omega_1$ and $\Omega_2$. There are several ways the coupling can be implemented. We chose a method in which the overlap extends on roughly one lattice site. The position of this site is indexed by $i=i_{\mathrm{int}}$ (see Figure~\ref{fig:interface}).

As a conclusion, the FD domain requires the knowledge of the values for $u_{i,j}$ for $i=i_{\mathrm{int}}$ and $j=1\cdots j_{\mathrm{max}}$, and the values for $v_{i,j}$ for $i=i_{\mathrm{int}}+1$ and $j=0\cdots j_{\mathrm{max}}$. Those values are easily obtained from the LB field that offers a natural access to the macroscopic variables~(\ref{eq:macroscopics}).

The LB domain on the other hand requires the knowledge of the $9$ values $f_{k;i,j}$ at $i=i_{\mathrm{int}}-1$ and $j=0\cdots j_{\mathrm{max}}$. Those values are more difficult to get at. Clearly, the LB values, offering a description of the fluid at a molecular level, contain more information than the FD values. In the next section, we present a first-order expansion of the $f^{(k)}$. It will show that they depend both of the values of the macroscopic variables and their gradients. Furthermore, the results of the first-order expansion will serve as a dictionary to convert from FD values to LB values.

\subsection{First-order expansion of the LB equation}\label{coupling:fo}
The first-order expansion of the LB dynamics is based on three approximations:
\begin{enumerate}
\item The streaming operator in equation~(\ref{eq:lbgk}) is replaced by a first-order time and space series:
\begin{equation}\label{eq:first-order}
f_k(\vec r+\vec v_i\delta t,t+\delta t)-f_k(\vec r,t)\approx\delta t\,\partial_t f_k(\vec r,t)+\delta t\left(v_{i\alpha}\partial_\alpha\right)f_k(\vec r,t)
\end{equation}
\item The values $f_k$ are split up into an equilibrium an non-equilibrium part. The spatial and temporal derivatives of the non-equilibrium part are neglected:
\begin{equation}\label{eq:feq-approximation}
f_k:=f_k^\mathrm{(eq)}+f_k^\mathrm{(neq)}\quad\textrm{and}\quad\partial_t f_k^\mathrm{(neq)}\approx0\textrm{,}\quad\partial_\alpha f_k^\mathrm{(neq)}\approx0.
\end{equation}
\item All second order velocity terms are neglected in the equilibrium distribution:
\begin{equation}\label{eq:truncated-feq}
f_k^\mathrm{(eq)}\approx a\,\rho+b\,\rho\frac{v_{k\alpha}u_\alpha}{v^2}
\end{equation}
\end{enumerate}
These approximations are consistent with the point of view taken by a first order Chapman-Enskog expansion (see for example~\cite{BC-livre}). In the following, we simplify the notation and replace the approximation sign ($\approx$) by a straight equality sign.

We start by expressing the conservation formulae~(\ref{eq:lb-mass-conservation})~and(\ref{eq:lb-momentum-conservation}) at the first order. From~(\ref{eq:lb-mass-conservation}) we obtain
\begin{equation}
\sum_k m_k\left(\partial_t f_k+v_{k\alpha}\partial_\alpha f_k\right)=0\quad\Rightarrow\quad\frac{\partial\rho}{\partial t}+\mathrm{div}(\rho\vec u)=0.\label{eq:mass-conservation}
\end{equation}
We consider the case of an incompressible fluid, thus
\begin{equation}\label{eq:lb-incompressibilite}
\frac{\partial\rho}{\partial t}=\mathrm{div}(\rho\vec u) =0.
\end{equation}
Momentum conservation~(\ref{eq:lb-momentum-conservation}) gives
\begin{equation}
\sum_k m_k\left(\delta t\,\partial_t v_{k\alpha}f_k+\delta r\,v_{k\alpha}v_{k\beta}f_k\right)=0\quad\Rightarrow\quad\partial_t(\rho u_\alpha)+\partial_\beta\Pi_{\alpha\beta}=0.\label{eq:momentum-conservation}
\end{equation}
This relation expresses the same physical content as the Navier-Stokes equations~(\ref{eq:navier-stokes}). We have introduced the stress tensor, defined as $\Pi_{\alpha\beta}=\sum_k m_k v_{k\alpha}v_{k\beta}f_k$. A second order analysis of the stress tensor enables to verify the equivalence between~(\ref{eq:navier-stokes}) and~(\ref{eq:momentum-conservation}), but this point is out of the scope of the present development.

With those relations, we are ready to consider the expansion of the LB dynamics~(\ref{eq:lbgk}). Using~(\ref{eq:first-order}), we find
\begin{equation}\label{eq:develop-start}
-\omega f_k^\mathrm{(neq)}+\frac{b\,\delta t}{v^2}\vec F\cdot\vec v_k = \delta t\,\partial_t f_k+\delta r(c_{k\alpha}\partial_\alpha)f_k\stackrel{\mathrm{(\ref{eq:feq-approximation})}}{=}\delta t\,\partial_t f_k^\mathrm{(eq)}+\delta r(c_{i\alpha}\partial_\alpha)f_k^\mathrm{(eq)}.
\end{equation}
The time-derivative of the equilibrium term is further expanded:
\begin{equation}
\partial_t f_k^\mathrm{(eq)}=\partial_\rho f_k^\mathrm{(eq)}\partial_t\rho+\partial_{\rho u_\alpha}f_k^\mathrm{(eq)}\partial_t(\rho u_\alpha)\stackrel{\mathrm{(\ref{eq:lb-incompressibilite},\ref{eq:momentum-conservation})}}{=}\frac{b\,v_{k\alpha}}{v^2}\left(-\partial_\beta \Pi_{\alpha\beta}+\delta t F_\alpha\right).
\end{equation}
From the isotropy relations~(\ref{eq:isotropy2},\ref{eq:isotropy3}), we know that
\begin{equation}
\partial_\beta\Pi_{\alpha\beta} = \frac{a}{b}v^2\partial_\alpha\rho.
\end{equation}
Thus,
\begin{equation}
\partial_t f_k^\mathrm{(eq)}=\frac{b\,v_{k\alpha}}{v^2}(-\partial_\beta\Pi_{\alpha\beta}+\delta t\,F_\alpha) = -a\,v_{k\alpha}\partial_\alpha\rho+\frac{b\,v_{k\alpha}}{v^2}\delta t\,F_\alpha.
\end{equation}
Plugging into~(\ref{eq:develop-start}) gives 
\begin{eqnarray}
f_k^\mathrm{(neq)} &=& \frac{\delta t}{\omega}\left(a\,v_{k\alpha}\partial_\alpha\rho-v_{k\alpha}\partial_\alpha\left(a\,\rho+b\,\rho\frac{v_{k\alpha}u_\alpha}{v^2}\right)\right)\nonumber\\
&=&-\delta t \frac{b}{\omega}\frac{v_{k\alpha}v_{k\beta}}{v^2}\partial_\alpha(\rho\,u_\beta)\label{eq:neq-terms}
\end{eqnarray}

This equation is the final lead in the chain to convert from the macroscopic variables to the set of LB variables. In order to summarize our findings, we remember that the $f_k$ are split up into an equilibrium and non-equilibrium part: $f_k=f_k^\mathrm{(eq)}+f_k^\mathrm{(neq)}$, where $f_k^\mathrm{(eq)} = f_k^\mathrm{(eq)}(\rho,u_\alpha)$ is obtained from equation~(\ref{eq:f_eq}), and $f_k^\mathrm{(neq)} = f_k^\mathrm{(neq)}(\rho,\partial_\alpha u_\beta)$ is approximated with the help of equation~(\ref{eq:neq-terms}).

For some values of the index $k$, the relation~(\ref{eq:neq-terms}) can be further simplified by imposing the continuity equation~(\ref{eq:continuity}). Table~\ref{table:dictionary} explicits the value of $f_k^\mathrm{(neq)}$ for every $k$.
\begin{table}
\begin{center}
\caption{Dictionary for the conversion from macroscopic variables to the non-equilibrium LB term.} \label{table:dictionary}

$t_k = -\omega/(\delta t\,b\,\rho)f_k^\mathrm{(neq)}$

\begin{tabular}{||l|c||l|c||l|c||}
\hline
$k$ & $t_k$ & $k$ & $t_k$ & $k$ & $t_k$\\
\hline\hline
$0$ & $0$ & $3$ & $\partial_x u_x$ & $6$ & $\partial_x u_y - \partial_y u_x$\\\hline
$1$ & $\partial_x u_x$ & $4$ & $\partial_y u_y$ & $7$ & $\partial_x u_y+\partial_y u_x$\\\hline
$2$ & $\partial_y u_y$ & $5$ & $\partial_x u_y + \partial_y u_x$ & $8$ & $-\partial_x u_y-\partial_y u_x$\\\hline
\end{tabular}

\end{center}
\end{table}

\subsection{Coupling the FD values}\label{coupling:FD}
Coupling the boundaries of the FD field to the LB field is a simple exercise. We have seen in section~\ref{models:fd} that the values of the pressure need not be coupled. The two components of the velocity field are computed in a straightforward manner from relations~(\ref{eq:macroscopics}). Because the FD and the LB variables are not defined at the same point of a lattice cell, we need to adjust the values by a linear interpolation. This leads to the following relations:
\begin{eqnarray}
u_{{i_\mathrm{int}},j}^\mathrm{FD} &=& 1/2\left(u_{{i_\mathrm{int}},j}^\mathrm{LB}+u_{{i_\mathrm{int}},j-1}^\mathrm{LB}\right)\\
v_{{i_\mathrm{int}}+1,j}^\mathrm{FD} &=& 1/2\left(v_{{i_\mathrm{int}}+1,j}^\mathrm{FD}+v_{{i_\mathrm{int}},j}^\mathrm{FD}\right)
\end{eqnarray}

\subsection{Coupling the LB values}\label{coupling:LB}
For the computation of the equilibrium terms $f_k^\mathrm{(eq)}$, we need to obtain the macroscopic variables $u$, $v$ and $p$ from the FD field. This is done by linear interpolation:
\begin{eqnarray}
u_{{i_\mathrm{int}}-1,j}^\mathrm{LB} &=& 1/2(u_{{i_\mathrm{int}}-1,j}^\mathrm{FD}+u_{{i_\mathrm{int}}-1,j+1}^\mathrm{FD})\\
v_{{i_\mathrm{int}}-1,j}^\mathrm{LB} &=& 1/2(v_{{i_\mathrm{int}}-1,j}^\mathrm{FD}+v_{{i_\mathrm{int}}-1,j}^\mathrm{FD})\\
p_{{i_\mathrm{int}}-1,j}^\mathrm{LB} &=& 1/4(p_{{i_\mathrm{int}},j+1}^\mathrm{FD}+p_{{i_\mathrm{int}}-1,j+1}^\mathrm{FD}+p_{{i_\mathrm{int}}-1,j}^\mathrm{FD}+p_{{i_\mathrm{int}},j+1}^\mathrm{FD})
\end{eqnarray}
A certain difficulty consists in relating the pressure field from the FD simulation to the density field of the LB simulation. Indeed, both fields contain a constant additive term which is {\it a priori} unknown. Expressing this term through an offset $p_\mathrm{ofs}$ of the pressure field, one has the following situation:
\begin{equation}
\frac{c_s^2\rho^\mathrm{(LB)}}{\rho_0^\mathrm{(LB)}} = p^\mathrm{(FD)}-p_\mathrm{ofs}
\end{equation}
We chose to fix this constant by averaging the FD pressure and the LB density on the interface between $\Omega_1$ and $\Omega_2$
\begin{equation}
\overline{\rho} = 1/h \int_{\partial\Omega_1\bigcap\partial\Omega_2}dr\,\rho\quad\textrm{and}\quad\overline{p} = 1/h \int_{\partial\Omega_1\bigcap\partial\Omega_2}dr\,p
\end{equation}
and claiming that the density average is constant on the interface: $\overline{\rho}=\rho_0$. This leads to
\begin{equation}
\rho = \rho_0\left(\frac{p-\overline{p}}{c_s^2}+1\right).
\end{equation}
It is not clear if this way of doing is optimal. Another solution might consist in solving locally the Poiseuille equation on the boundary of the LB field. This would enable to compute the values for $\rho$ from LB variables only.

The non-equilibrium terms $f_k^\mathrm{(neq)}$ are based on the gradients of the velocities. Figure~\ref{fig:interface} shows that two of those gradients can be approximated by a centered difference of half the mesh width:
\begin{eqnarray}
\partial_y u_{{i_\mathrm{int}}-1,j}^\mathrm{LB} &=& 1/\delta r(u_{{i_\mathrm{int}}-1,j+1}^\mathrm{FD}-u_{{i_\mathrm{int}}-1,j}^\mathrm{FD})\\
\partial_x v_{{i_\mathrm{int}}-1,j}^\mathrm{LB} &=& 1/\delta r(v_{{i_\mathrm{int}},j}^\mathrm{FD}-v_{{i_\mathrm{int}},j-1}^\mathrm{FD})
\end{eqnarray}
One gradient needs to be calculated as a centered difference of mesh width, based on interpolated values of the velocity:
\begin{equation}
\partial_y v_{{i_\mathrm{int}}-1,j}^\mathrm{LB} = \frac{1}{4\delta x}(v_{{i_\mathrm{int}},j+1}^\mathrm{FD}+v_{{i_\mathrm{int}}-1,j+1}^\mathrm{FD}-(v_{{i_\mathrm{int}},j-1}^\mathrm{FD}+v_{{i_\mathrm{int}}-1,j-1}^\mathrm{FD})).
\end{equation}
There are not enough grid points at hand for computing the fourth gradient at the same level of precision. We are luckily saved by the continuity equation~(\ref{eq:continuity}) which delivers the requested value:
\begin{equation}
\partial_x u_{{i_\mathrm{int}}-1,j}^\mathrm{LB} = -\partial_y v_{{i_\mathrm{int}}-1,j}^\mathrm{LB}
\end{equation}

Now that all missing variables have been computed, we take a step back and discuss the overall algorithm of a LB iteration step. For the purpose of this discussion, the dynamics~(\ref{eq:lbgk}) are split into two steps. The first, the collision step, handles the computation of the equilibrium distribution and maps the ``incoming particle stream'' $f_k^\mathrm{(in)}$ onto the ``outgoing particle stream'' $f_k^\mathrm{(out)}$. It is followed by a streaming step that transports the particles by a value of $\delta t$ in time and $\vec v_k\delta t$ in space. The details of an iteration step are as follows:
\begin{enumerate}
\item On bulk nodes, $\rho(t)$ and $\vec u(t)$ are computed from the incoming particle densities $f_k^\mathrm{(in)}(t)$. On boundary nodes, all the values $\rho(t)$, $\vec u(t)$ and $f_k^\mathrm{(in)}(t)$ are obtained from the variables of the FD field at time $t$.
\item All nodes, bulk and boundaries, perform the collision step: $f_k^\mathrm{(out)}(\vec r,t) = (1-\omega)f_k^\mathrm{(in)}(\vec r,t)+\omega f_k^\mathrm{(eq)}(\vec r,t)$.
\item The bulk nodes perform the streaming step: $f_k^\mathrm{(in)}(\vec r+\vec v_k\delta t,t+\delta t) = f_k^\mathrm{(out)}(\vec r,t)$, for all $\vec r$ such that $\vec r+v_k\delta t$ lies on a bulk node.
\end{enumerate}
Alternatively, it is possible to extend the streaming step to boundary nodes for those values of $f_k^\mathrm{(in)}$ that are incoming from the bulk of the LB simulation. They are kept unchanged, unlike the remaining set of $f_k^\mathrm{(in)}$ and the macroscopic variables $\rho$ and $\vec u$ that are provided by the FD field. In our simulations, this procedure seemed to produce results equivalent to those of the proposed algorithm.

\section{Validation on a Poiseuille flow}\label{poiseuille}
We propose a validation of our coupling algorithm on the simulation of a Poiseuille flow. This is a stationary flow in a channel of infinite length with no-slip boundaries. The boundaries extend horizontally at a height $y=0$ and $y=L$. The fluid velocity is strictly horizontal and does not depend on the $x$ position: $v=0;\quad \partial_x u=0$. The analytical solution of the Navier-Stokes equations~(\ref{eq:navier-stokes}) for this problem is known and predicts a parabolic velocity profile:
\begin{equation}\label{eq:parabola}
u(y) = \frac{1}{2\nu}C(Ly-y^2).
\end{equation}
The constant $C$ can be related to the body force ($C=f_x$) or to the pressure gradient ($C=-\partial_x p$), depending on how the fluid is driven.

In our example, the fluid is driven by a body force. The left and the right borders implement periodic boundary conditions in order to simulate a channel of infinite length. Special care must be taken on specifying this boundary in the FD model. Indeed, during the simulation it tends to build up a pressure gradient that must be eliminated by imposing a constant value of the pressure on the left and right boundary.

The simulation is performed on a grid of size $i_\mathrm{max}=3$ and $j_\mathrm{max}=50$. The physical channel widht is set to $L=1$ and the body force has the value $F_x=0.01$. The numeric values $u_\mathrm{sim}$ are compared to the analytic solution $u_\mathrm{ana}$ from equation~(\ref{eq:parabola}) by means of the overall error (the indices of the formula refer to the LB convention introduced on Figure~\ref{cell}):
\begin{equation}\label{eq:error}
\epsilon =
  \frac
    {
      \sqrt{ \sum_{j=0}^{j_\mathrm{max}}(u_\mathrm{sim}(\cdot,j)-u_\mathrm{ana}(\cdot,j))^2 }
    }
    {
      \sqrt{ \sum_{j=0}^{j_\mathrm{max}}u_\mathrm{ana}(\cdot,j)^2 }
    }
\end{equation}

\begin{figure}
\begin{center}
\caption{The computational grid for the simulation of a Poiseuille flow is partitioned into three subdomains $\Omega_1$, $\Omega_2$ and $\Omega_3$. The FD scheme is used on the boundary domains $\Omega_1$ and $\Omega_3$, and the LB scheme on the bulk domain $\Omega_2$.}\label{fig:poiseuille}

\begin{pspicture}(9.5,4.5)
\psset{unit=0.4}
\psline[linewidth=0.1,linestyle=dotted](0,2)(16,2)
\psframe[fillstyle=solid,fillcolor=lightgray,linestyle=none](0,0)(16,2)
\psline[linewidth=0.1,linestyle=dotted](0,8)(16,8)
\psframe[fillstyle=solid,fillcolor=lightgray,linestyle=none](0,8)(16,10)
\psline[linewidth=0.1](0,0)(16,0)
\psline[linewidth=0.1](0,10)(16,10)
\psline[linewidth=0.1,linestyle=dashed](0,0)(0,10)
\psline[linewidth=0.1,linestyle=dashed](16,0)(16,10)
\psline[linewidth=0.1]{<->}(18,0)(18,10)
\psline[linewidth=0.1]{->}(20,0)(23,0)
\psline[linewidth=0.1]{->}(20,0)(20,3)
\rput[bl](23.1,0.1){$x$}
\rput[bl](20.1,3.1){$y$}
\rput[bl](18.3,7.3){\large$L=1$}
\rput[bl](18.3,5.3){\large$j_\mathrm{max}=50$}

\psline[linewidth=0.05]{->}(1,5)(4,5)
\rput[b](2.5,5.2){\large$\vec u$}

\rput[bl](6,8.1){\large$\Omega_3$(FD)}
\rput[bl](6,4.5){\large$\Omega_2$(LB)}
\rput[bl](6,0.1){\large$\Omega_1$(FD)}

\rput[bl](11,8.1){\large$j_\mathrm{max}^{(3)}=3$}
\rput[bl](11,4.5){\large$j_\mathrm{max}^{(2)}=46$}
\rput[bl](11,0.1){\large$j_\mathrm{max}^{(1)}=3$}

\end{pspicture}
\end{center}
\end{figure}
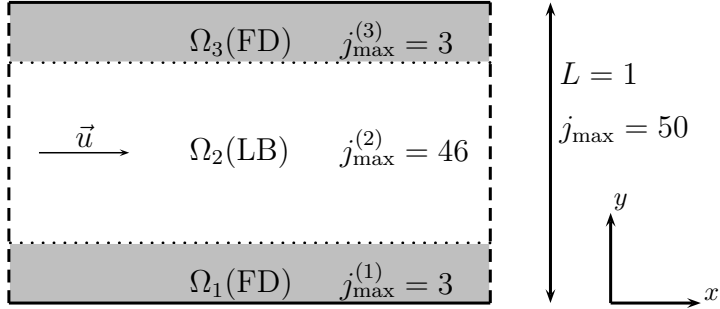

We have run three simulations that can be compared among each other. The first simulation implements a pure LB model with bounce-back boundaries, the second simulation a pure FD model, and the third simulation is a FD--LB hybrid. In the first case, the no-slip property of the walls is implemented by a boundary condition known under the name of bounce-back (see {\it e.g.}~\cite{BC-livre}). In the third case, the top and bottom strips of size $j_\mathrm{max}^\prime=3$ are computed by the FD model and the bulk domain by the LB model (see Figure~\ref{fig:poiseuille}).

\begin{figure}
\begin{center}
\caption{Simulation of a body-force driven Poiseuille flow with (1) a LB model, (2) a FD model and (3) a FD-LB hybrid (see Figure~\ref{fig:poiseuille}). The curves show the time-evolution of the error, compared to the analytical solution of the Poiseuille flow.}\label{fig:convergence}
\includegraphics[width=8cm]{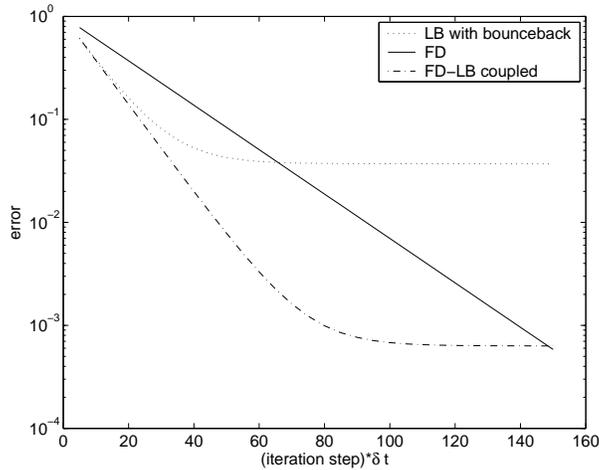}
\end{center}
\end{figure}

A remarkable result of the simulations is that the FD-only model reaches the analytical solution at the machine level of precision ($10^{-15}$). Although there exist LB boundary conditions which obtain the same result~(\cite{inamuro:95}), their implementation is less natural and straightforward than the one of the FD model. We further remark that the LB model has a faster convergence (in terms of iteration steps) than the FD model. The stationary velocity field of the LB simulation is dock distinct from the analytical prediction, due to the limited precision of the boundary condition that is known to be of first order. The hybrid simulation shows the expected convergence speed of the LB model and an error due to the limited precision of the coupling. However, the error is two orders of magnitude smaller than the one due to the bounce-back boundaries of the LB-only simulation. The results of the simulations are shown on Figure~\ref{fig:convergence}.

\begin{figure}
\begin{center}
\caption{Error of a FD-LB hybrid Poiseuille flow simulation as a function of the grid resolution. A log-log plot shows the coupling of the velocity field to be of second order.}\label{fig:order-of-the-coupling}
\includegraphics[width=8cm]{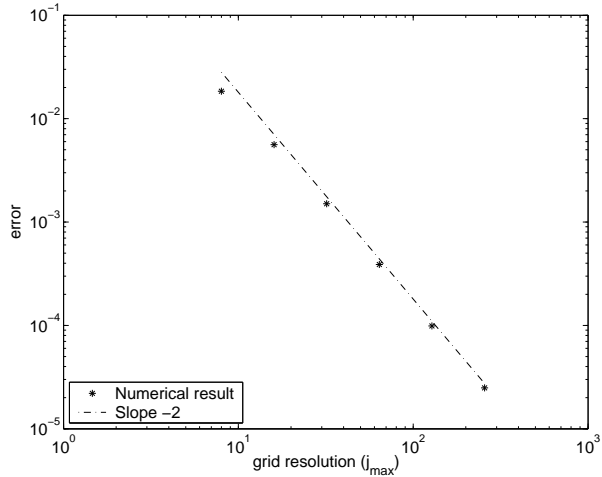}
\end{center}
\end{figure}

The order of precision of the coupling can be estimated by varying the grid resolution ($j_\mathrm{max}$) while keeping the physical quantities($L$, $F_x$) constant. Figure~\ref{fig:order-of-the-coupling} plots the error of the stationary velocity field as a function of the grid resolution. It appears clearly that the coupling acts like a second-order boundary condition for the velocity field. No conclusion can be taken concerning the coupling of the pressure field, because the latter is constant in a Poiseuille flow.

\section{Conclusion}\label{conclusion}
In this work, a LB scheme for 2-D incompressible fluid flows is spatially coupled to a FD scheme on a computational domain partitioned in two regions. We present a way to relate the LB distribution functions $f_k$ with the classical physical quantities and their derivatives. Two particular FD and LB schemes are introduced, and a complete coupling algorithm between the two is proposed. At the interface, the LB and FD are connected so as to preserve continuity of the physical quantities. The connection between the $f_k$ variables and the standard macroscopic physical quantities is obtained through the analysis of a first-order truncated series around the local equilibrium. The equilibrium part of $f_k$ is related to the macroscopic quantities and the non-equilibrium part to the gradients thereof. Our coupling methodology is indeed an approximation since we neglect higher-order derivatives in the nonequilibrium distributions. A validation was performed by simulating a Poiseuille flow with FD boundary strips and LB bulk and comparing it with an analytic solution. The simulation shows that in this case, the coupling of the velocity field is of second order in the grid resolution.

We consider the work on the LB-FD coupling to be interesting by its own means, as it expresses the conceptual differences between the approaches of those models. In particular, it might be inspiring in formulating new kinds of boundary conditions for either model. In this sense, the FD model can take profit of the physical point of view taken in the LB approach, whereas LB boundary condition can be inspired by the strict mathematical formulations of the FD boundaries.

The value of our hybrid model in practice needs still to be shown, but we are confident that it will prove itself useful in a large class of scientific and engineering problems. We plan to provide first sample applications by implementing drag force experiments with LB obstacles and FD domain boundaries.


\begin{thebibliography}{1}

\bibitem{wittwer:03}
Peter Wittwer, Sebastian {B\"onisch}, and Vincent Heuveline.
\newblock Adaptive boundary conditions for exterior flow problems, 2003.
\newblock Preprint:
  http://pcc2341f.unige.ch/publications/paper029/paper029.pdf.

\bibitem{davide:04}
Paul Albuquerque, Davide Alemani, Bastien Chopard, and Pierre Leone.
\newblock Coupling a lattice boltzmann and a finite difference scheme.
\newblock In {\em ICCS 2004}. Krakow, Poland, 2004.

\bibitem{succi-book:01}
Sauro Succi.
\newblock {\em The Lattice Boltzmann Equation, For Fluid Dynamics and Beyond}.
\newblock Oxford University Press, 2001.

\bibitem{BC-livre}
B.~Chopard and M.~Droz.
\newblock {\em Cellular Automata Modeling of Physical Systems}.
\newblock Cambridge University Press, 1998.

\bibitem{wolf-gladrow:00}
Dieter~A. Wolf-Gladrow.
\newblock {\em Lattice-Gas Cellular Automata and Lattice Boltzmann Models: an
  Introduction}.
\newblock Lecture Notes in Mathematics, 1725. Springer, Berlin, 2000.

\bibitem{griebel-book}
Michael Griebel, Thomas Dornseifer, and Tilman Neunhoffer.
\newblock {\em Numerical Simulation in Fluid Dynamics. A Practical
  Introduction}.
\newblock Siam, Philadelphia, 1998.

\bibitem{inamuro:95}
Takaji Inamuro, Masato Yoshino, and Fumimaru Ogino.
\newblock A non-slip boundary condition for lattice {B}oltzmann simulations.
\newblock {\em Phys. Fluids}, 7(12):2928--2930, 1995.

\end{thebibliography}
\end{document}